# Do wind-generated waves under steady forcing propagate primarily in the downwind direction?

Paul A. Hwang, David W. Wang

*Oceanography Division, Naval Research Laboratory, Stennis Space Center, MS*

James Yungel, Robert N. Swift

*EG&G, N159, WFF, Wallops Island, VA*

William B. Krabill

*NASA/GFSC/WFF, Wallops Island, VA*

## Abstract

Measuring the directional distribution of ocean waves is a difficult task in ocean wave research. For many decades, wind-generated waves are assumed to propagate primarily in the wind direction. The concept is applied either implicitly or explicitly in the formulation of the wind input function and the design of the wave directional distribution function. Photographs of the ocean surface wave fields under steady wind forcing, however, display very different directional properties. In the near shore region, the surface undulations are dominantly crosswind. As the fetch increases, the geometry of surface waves is crosshatched, indicating two dominant wave trains crossing each other. This pattern of wave propagation in off-wind directions is repeatedly photographed in different locations under different offshore (toward) wind events. The wavelengths of these airborne photographs are in the decameter length scale. Shorter waves with wavelengths ranging from a few centimeters to a few decimeters also show crosshatched patterns, based on pictures taken from land and ship. In the crosshatched cases, most pictures show that the wind vector bisecting the diamond shape of the surface wave geometry. The observed wave geometries and their fetch evolution expose a major flaw in our fundamental assumption (implicitly or explicitly) that "wind-generated waves propagate primarily in the wind direction." The assumption has been taken for granted for many decades in applications including basic research of wave dynamics and the design of directional wave spectral model for calculations of the ocean surface roughness and the force, energy, and momentum of a wave field.

## 1. Introduction

It is generally accepted as a fact that wind generates waves, and that under steady wind forcing, wind-generated waves propagate in the wind direction. This concept is built-in implicitly or explicitly in the design of the directional distribution function. For example, the distribution functions of Mitsuyasu et al. (1975) and Hasselmann et al. (1980) assume a cosine function raised to the $2s$ power, $\cos^{2s}[(\theta-\theta_m)/2]$, where $\theta$ is the angle measured from the wind direction, $\theta_m$ is the dominant direction of wave propagation, and $s$ is a parameter related to the directional beam width. Donelan et al. (1985) suggest a hyperbolic secant squared function, $\text{sech}^2[b(\theta-\theta_m)]$, where $b$ is related to the directional beamwidth. While details of various functions differ, the symmetric functional forms stipulate that the dominant directions of all spectral components to be in the wind direction. These directional functions are used in virtually all wave spectral models for scientific and engineering applications. In a similar fashion, in the absence of measurements on the directional dependence of the input function of wind-wave generation, a cosine directional dependence is assumed in the energy or action conservation equation of ocean wave dynamics. Such assumption has rarely been questioned mainly because of our deep-rooted trust that "wind-generated waves propagate primarily in the wind direction."

In Section 2, we show photographs of ocean waves generated by steady offshore wind events. Surprisingly, the directional distributions of the fetch-limited wave fields are quite different from that expected of unimodal directional propagation. In particular, in the nearshore region the propagation direction of surface waves is crosswind. At a longer fetch, crosshatched patterns are frequently observed. Section 3 presents a qualitative discussion of the possible mechanism of off-wind propagation of surface waves. It is suggested that the observed wave patterns and their fetch evolution appear to reflect the character of resonant propagation between surface waves and the forcing wind field. A conceptual model of the wind input function considering the resonance effect is proposed. Section 4 presents a summary.

## 2. Photographs of wind-generated surface wave patterns

Plate 1 shows two images of the wave conditions at different fetches along a flight track in the Gulf of Mexico. (The flight tracks presented in this paper are all along or against the wind direction and the aircraft flies repeat linear or racetrack pattern. The wind directions in the photographs are mostly in the left/right orientation but the details of the photographic angles are not recorded). The waves are generated by a steady offshore wind following a cold front passing through the region. The coastline is simple, quasi-linear and almost perpendicular to the wind vector. The maximum fetch coverage of the flight is 42 km; more details of the experiment are described in Hwang et al. (2019). The top image (Plate 1a) is in the near shore region immediately connected to the coastline, which is visible in the middle of the image. The bottom image (Plate 1b) is taken near the far end of the flight track. In both images, it is difficult to argue that waves are traveling in the downwind direction. In fact, the propagation direction of the surface undulations in the first image is perpendicular to wind. The crosshatched pattern in the second image indicates that two wave systems of about equal intensity crossing each other at a large angle. It is quite obvious that these wave patterns do not fit the model of conventional unimodal directional distribution functions such as those described in the Introduction section.

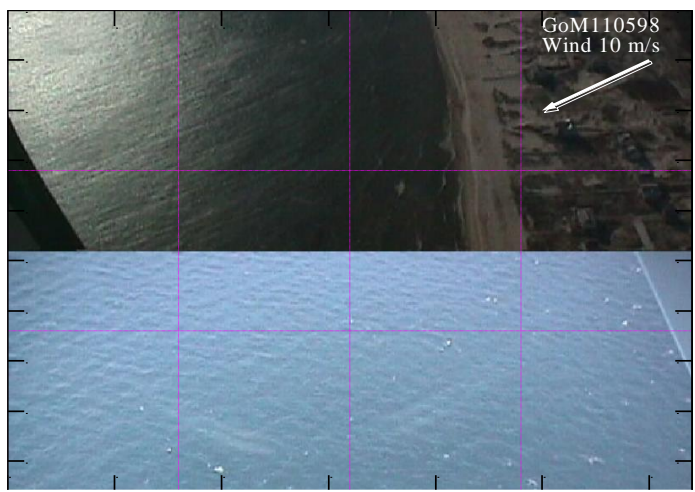

Plate 1. Two photographs of surface waves taken along a flight track that extends from shore to 42 km offshore in the Gulf of Mexico. The top image is taken at a nearshore location. The bottom image is taken near the far end of the flight track. Steady offshore wind condition prevails during the flight.

To be certain that what's shown in Plate 1 is not an abnormal occurrence, many more airborne photographs of wind–generated waves have been acquired under steady offshore wind conditions. The patterns of crosswind undulations in the near shore region and offwind propagation in the offshore regions are again observed. For example, Plate 2a is taken near the coastal region at Hatteras Bight, North Carolina. The surface undulations are dominantly linear features perpendicular to the coastline. The propagating direction is perpendicular to the wind vector instead of downwind. The wave image in Plate 2b is taken in the Atlantic Ocean south of Hatteras Bight on another offshore wind event. The streaks of foam lines unveil the wind direction, which is significantly different from the propagation directions of the dominant wave components. Plate 3a is taken near the coastal region of Assateague Island, Virginia. The wind is steady offshore westerly. The wave pattern is again linear streaks parallel to wind. Plate 3b is taken at about 30 km from the coast. Two wave systems propagating at oblique angles can be detected from the surface topography. These wave images certainly disagree with the assumption that wind-generated waves propagate in the wind direction. Instead, they suggest bimodal directional propagation of surface waves in fetch-limited growth conditions. The wave propagation direction varies from ±90° (crosswind) in the near shore region to approximately ±45° at 30 to 40 km offshore for 10 to 12 m/s wind speeds. The resulting surface pattern is crosshatched or diamond shaped.

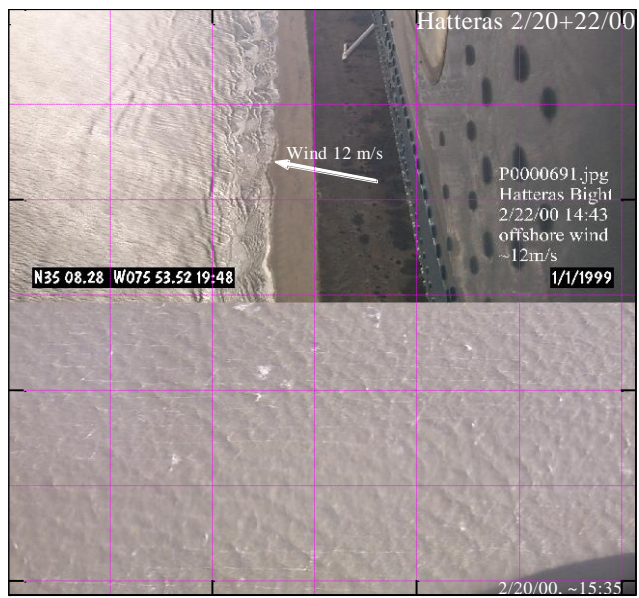

Plate 2. Two more photographs of surface wave patterns under steady offshore-directed wind forcing near the Hatteras Bight. Similar to the case encountered in the Gulf of Mexico displayed in Plat 1, in the near shore region (top), linear undulations with wave fronts perpendicular to the wind direction are the dominant feature. In the offshore region, crosshatched wave patterns are common (bottom).

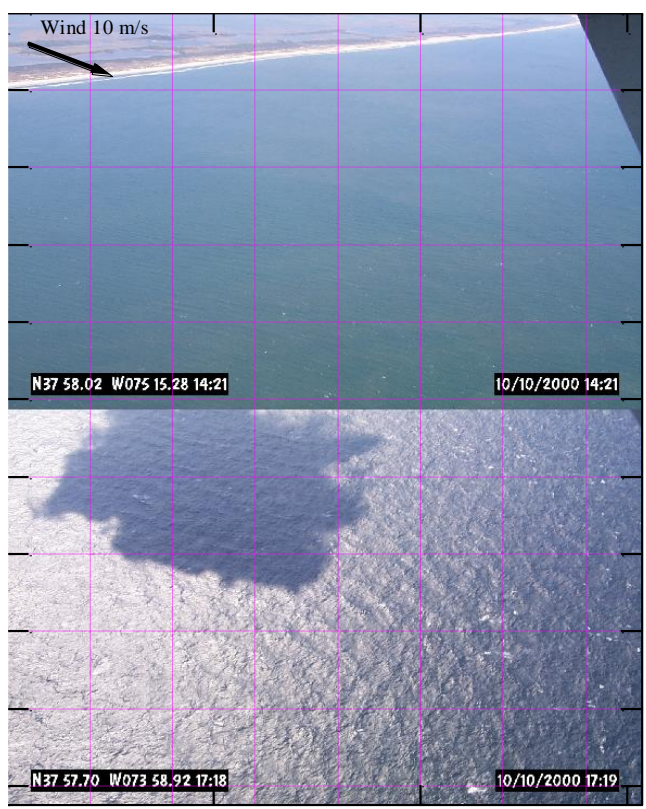

Plate 3. Nearshore crosswind wave pattern (top) and offshore bimodal directional pattern (bottom) of surface waves obtained near Assateague Island, Virginia.

    The wavelengths in the airborne photographs are in the decameter range. Bimodal directional propagation is also found in small-scale waves. Plate 4 shows an example taken at a canal at Stennis Space Center (SSC), Mississippi. The canal is approximately 600 m long and 100 m wide. The wind direction is along the long axis of the canal. The location of the photograph is near the mid-length of the canal, and the camera is pointing crosswind toward the opposite shore. The dominant waves are obviously propagating at off-wind angles. The white haul of the stationary barge across the canal provides excellent lighting to reveal the diamond pattern of the wave geometry aligned symmetrically to the wind vector.

Plate 5 is a photograph of very small-scale waves taken in the Puget Sound, Washington, during a field test of a free-drifting scanning slope sensor buoy. For scale reference, the diameter of the white floatation column is 25 cm and the diameter of the structural members is 5 cm. The wavelength of the crosshatched waves is estimated to be less than 10 cm. These waves are generated by a gentle breeze.

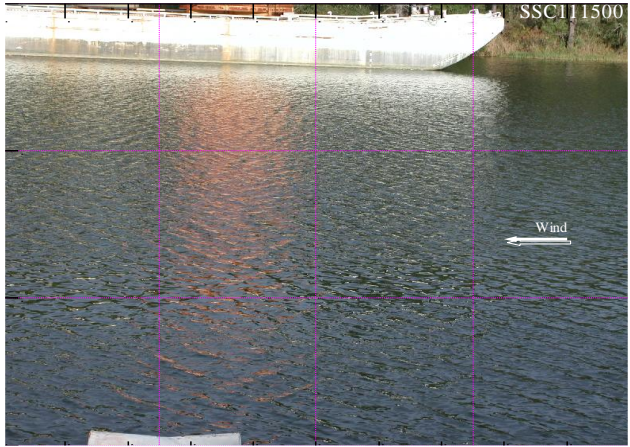

Plate 4. Bimodal directional propagation of short waves in a canal. The wavelengths are on the order of a few decimeters. The waves are produced by a steady wind coming from the right hand side of the picture.

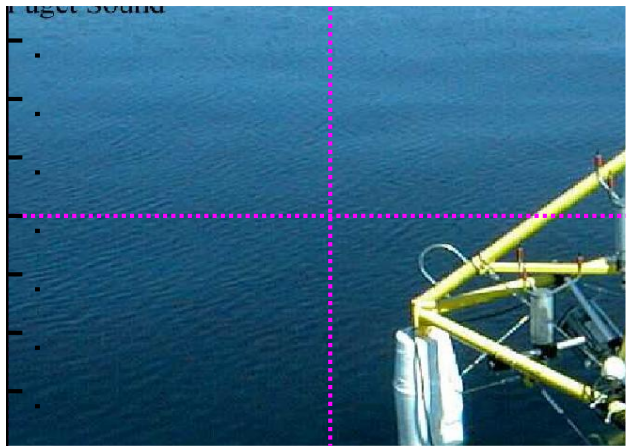

Plate 5. Bimodal directional propagation of short waves with wavelengths on the order of a few centimeters. The waves are produced by a gentle breeze in the Puget Sound.

Another example of the bimodal propagation patterns of small-scale waves is shown in Plate 6. This one is taken in a small bayou (approximately 5 m wide) in Louisiana. A light breeze is blowing from the right hand side of the picture and the camera is pointing crosswind toward the opposite side of the bayou. The lighting condition near the center of the picture is just right for capturing the beautiful diamond pattern of the short waves. The wavelength is also in the sub-decimeter scale.

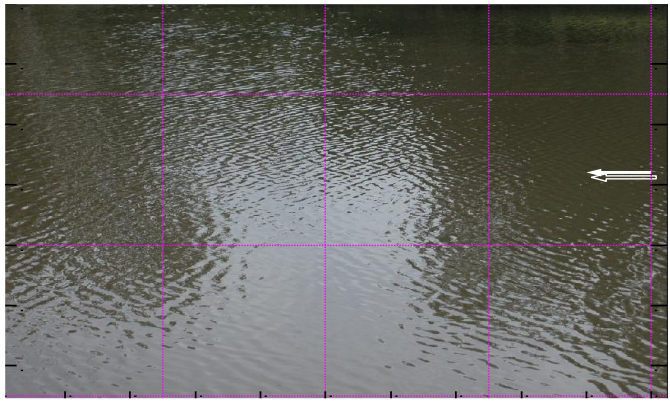

Plate 6. Bimodal directional propagation of short waves with wavelengths on the order of a few centimeters. The waves are produced by a gentle breeze in a Louisiana bayou.

## 3. Discussions

The directional patterns of wind-generated surface waves presented in the last section clearly show directional bimodality of the dominant wave component. This directionally bimodal wave pattern deviates from our intuition that "wind-generated waves propagate in the wind direction." Instead, in the near shore region the propagation direction is crosswind (Plates 1-3 top panels). Far away from the coast, the surface geometry is predominantly crosshatched. The pictures of Plates 1-3 bottom panels, and 4 to 6 are especially instructive. They indicate that two wave trains of approximately equal amplitudes and wavelengths crossing each other. The propagation directions of the two wave trains are about equal with respect to the wind direction. In all cases of airborne photographs, measurements from nearby NDBC (National Data Buoy Center) buoys confirm that the wind conditions are steady offshore blowing. In the Gulf of Mexico case (November 5, 1998, 1900-2100 UTC), the wind velocity is 10 m/s from NNW (Hwang et al. 2019). In the Hatteras Bight cases, the wind velocity is 10 m/s from north on February 20, 2000 (1535 UTC) and 12 m/s from north on February 22, 2000 (1443 UTC). In the Assateague Island case, the wind velocity is 10 m/s from west on October 10, 2000 (1400-1800 UTC). The propagation of two symmetrical wave trains oblique to wind is also shown in the SSC canal case (Plate 4). The wind direction is steady from South (along the long axis of the canal). In the cases of small waves obtained in the Puget Sound and a Louisiana bayou, the wind condition is light breeze.

The sequence of directional variation from crosswind propagation in the near shore region to a large oblique angle at further downwind locations suggests that resonant propagation is an important feature of wind generated surface waves. Qualitatively, as wind impinges on the water surface, waves of many length scales are generated and they propagate in a broad range of directions. Because these young waves travel much slower than the wind, their growth is transient and only those wave components that are in resonant propagation with the wind field receive continuous enrichment from the wind. The propagation direction satisfying the resonant propagation condition is $U\cos\theta_r=C$, where $U$ is wind speed, $C$ is wave speed and $\theta_r$ is the resonant wave propagation angle measured from the wind direction. Phillips (1957) has suggested such mechanism of resonance wind wave generation several decades earlier. Kinsman (1965, p. 542) includes several pictures of surface waves illustrating the regular rhomboid structure of the wind-generated wave field expected from Phillips (1957) resonance theory of wind-generated waves.

More quantitative off-wind propagation of dominant waves has been documented in the airborne scanning radar wave measurements by Walsh et al. (1985, 1989). The radar data show one wind-generated wave system propagating at an angle deviating from the prevailing wind direction. The deviation angle decreases with increasing fetch, and eventually the wave system propagates in the wind direction at fetches greater than 120 to 150 km for 12 to 17 m/s winds. They attribute the off-wind propagation to slant fetch effect. Long et al. (1994) and Huang (1999) present reanalysis of the airborne scanning radar data. They show that the fetch-dependent angle between wave and wind directions is in agreement with the calculated angle satisfying the resonant propagation condition. The airborne scanning radar data only show one oblique wave system instead of two symmetric systems as predicted by the resonant propagation condition. Long et al. (1994) and Huang (1999) suggest that one of the two systems is suppressed by the background swell present at the time of data collection. For the frontal passage event in which the photographs of Plate 1 are taken, an airborne scanning lidar system is used to collect ocean surface topography data. The aircraft flies a racetrack pattern with two long legs aligned to the wind direction. The fetch coverage is from coast to 42 km offshore. Fig. 1 (left column) shows a sequence of the surface topography at 6 different fetches along the track. For each image panel, the wind is from the right hand side. The directional spectrum of each topographic image is displayed in the right panel next to the image. Continuous evolution of the spectrum clearly shows the trend of dominant wave component moving from perpendicular-to-wind toward along-wind as wave grows with increasing fetch. These airborne lidar measurements, together with the photographs presented in this paper show convincingly that bimodal directional propagation of two dominant wave systems is indeed a frequent occurrence of surface waves generated by steady winds. The range of the length scale of waves undergoing bimodal directional propagation is very broad, from several centimeters to several decameters are shown in the small collection of photographs here.

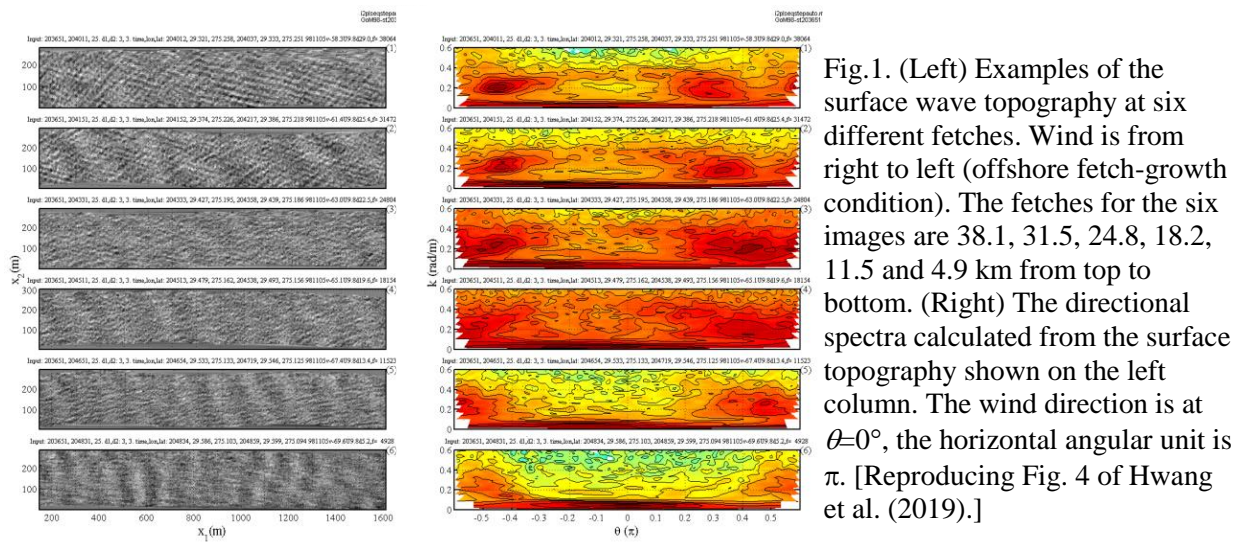

Fig.1. (Left) Examples of the surface wave topography at six different fetches. Wind is from right to left (offshore fetch-growth condition). The fetches for the six images are 38.1, 31.5, 24.8, 18.2, 11.5 and 4.9 km from top to bottom. (Right) The directional spectra calculated from the surface topography shown on the left column. The wind direction is at $\theta=0°$, the horizontal angular unit is $\pi$. [Reproducing Fig. 4 of Hwang et al. (2019).]

Extensive efforts have been devoted to the search of the true directional distribution of wind-generated ocean waves. One consensus derived from the earlier efforts is that "wind generates waves, and that under steady wind forcing, wave components of different length scales propagate primarily in the wind direction." The major difference of several established directional models is in the directional spreading of individual wave components. Over the years, directional distribution functions are assumed directionally unimodal, established principally on the firm belief that under steady wind forcing, wind-generated waves propagate in the wind direction. Photographs of wave fields generated by steady wind forcing do not support the notion of unimodal directional distribution. For long scale waves (wavelengths in the decameter range), airborne photographs provide large area coverage. Three sets of examples are shown in Plates 1 to 3. These images show that in the near shore region, the dominant surface geometry is composed of linear features with their normal pointing in the crosswind direction. As the fetch increases, crosshatched patterns develop, suggesting the presence of two wave systems crossing each other. The propagation directions of the two wave systems are not in the wind direction. The crosshatched wave pattern is also frequently observed in shorter scale waves. Plate 4 shows a picture taken in a canal about 100 m wide and 600 m long. The dominant wavelength of the crosshatched surface wave pattern is several decimeters. The wavelength in Plates 5 and 6, taken respectively in the Puget Sound and a Louisiana bayou, is only several centimeters long. The prevalence of the crosshatched surface wave patterns strongly suggests that our present concept of "wind-generated waves propagate in the wind direction" is not accurate. For fetch-limited waves that propagate slower than the wind speed, the dominant-scale waves propagate in off-wind directions to maintain propagation resonance. For waves reaching equilibrium condition, downwind propagation occurs for a narrow wavenumber band around the dominant wavelength. The wave components shorter than the dominant waves are also directionally bimodal due to nonlinear wave-wave interaction (Banner and Young, 1994; Hwang et al., 2000a-c; Hwang and Wang, 2001).

The observed bimodal directional patterns in the broad wavelength range of surface waves generated by steady wind also offer some hints on the wavenumber/directional dependence of the source function of wind-wave generation. A conceptual sketch is shown in Fig. 2, reflecting the bell-shaped (in 1D) or cone-shaped (in 2D) character of the response function of a resonant system. The general function form can be expressed as

$$R(k,\theta) = \frac{\mu'}{\left[1 - \left(\frac{\omega}{\omega_r}\right)^m + \varepsilon(k)\right]^n}, \quad (1)$$

where $\omega_r = k \bullet U$, represents the resonance condition of the wind-wave system, $k$ is the wavenumber vector with magnitude $k$ and direction $\theta$, $U$ is the reference wind velocity vector, $\omega$ is the angular frequency of the wave component, $\varepsilon$ is the dissipation function of the wave system, and $\mu'$ is an unspecified function related to the growth rate. Present formulations of the source function only contain the growth rate of the wind wave system and do not reflect the resonance nature of wind-wave generation.

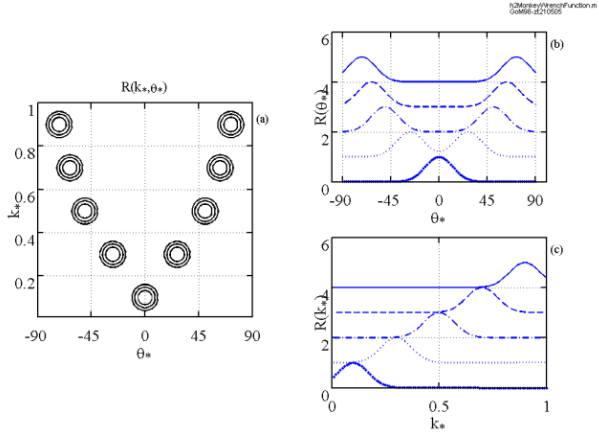

Fig. 2. A conceptual sketch of the source function reflecting the resonance mechanism of wind-wave generation. (a) 2D source function, (b) 1D directional function, and (c) 1D wavenumber dependence. Fetch increases from top to bottom. For clarity, the 1D functions in (b) and (c) at 5 fetches (increasing from top to bottom) are plotted with incremental offset.

It is noted that the resonance mechanism is inherent in both solutions of Phillips (1967) and Miles (1957, 1962) wind-wave generation mechanisms. The solution combining both mechanisms can be expressed as (Phillips, 1977; Massel, 1996)

$$\Psi(k,t) = \frac{\pi \Psi_a(k,\omega)}{\rho_w C^2} \frac{\sinh(\mu\omega t)}{\mu\omega}, \quad (2)$$

where $\Psi(k,t)$ is the directional wavenumber spectrum of ocean waves at time $t$, $\rho_w$ is the water density, $\mu$ is a growth rate parameter, $\omega$ is the angular frequency, and $\Psi_a(k,\omega)$ is the wavenumber-frequency spectrum of the atmospheric forcing function. The argument of $\Psi_a$ is defined by $\omega = k \bullet U$, indicating the resonant or selective nature of wind-wave generation. Presently, the studies of the growth rate of wind-generated surface waves generally address the second part on the right hand side of (2), $\sinh(\mu\omega t)/\mu\omega$. The atmospheric forcing term ($\Psi_a(k,\omega)$) that specifies the selective process ($\omega_r = k \bullet U$) and requires the knowledge of the full three-dimensional wavenumber-frequency spectrum of the atmospheric pressure field has not been fully resolved (O. M. Phillips, private communication, 2001). The conceptual model depicted in Fig. 2 is constructed with a bi-Gaussian function of the form

$$R(k,\theta,\omega) \sim \exp\left[-\left(\frac{k-k_r}{\sigma_k}\right)^2\right] \exp\left[-\left(\frac{\theta-\theta_r}{\sigma_\theta}\right)^2\right], \quad (3)$$

where $\sigma_k$ and $\sigma_\theta$ are indices of the "sharpness" of the resonance response function.

Using the continuous data of the directional wavenumber spectra derived from the airborne scanning lidar system (Fig. 1), the 2D rate of change of the wave system can be calculated from finite

differencing

$$\frac{\mu(k,\theta)}{\omega} = \frac{1}{N(k,\theta)} C_{gx} \frac{dN(k,\theta)}{dx}, \quad (4)$$

where $N$ is the action density function, and $C_{gx}$ is the downwind component of the wave group velocity. Using spectra calculated from consecutive segments of 1.6-km each, the resulting 2D growth rate function is shown in Fig. 3.

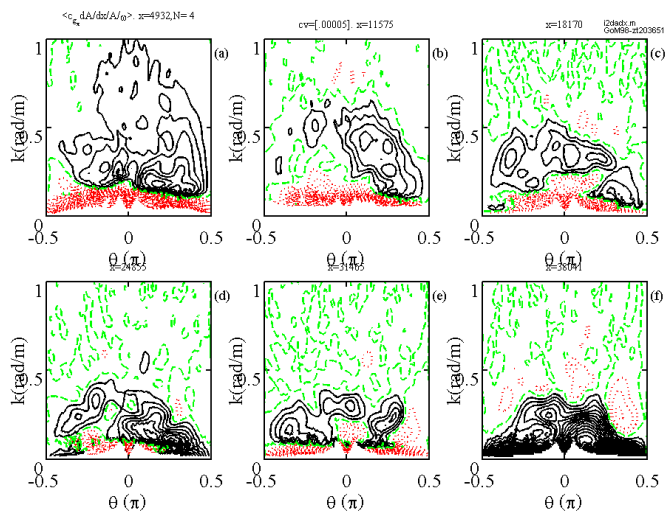

Fig. 3. The growth rate function derived from finite differencing the continuous data of ocean wave directional spectra, shown in Fig. 1, acquired by the airborne scanning lidar system. The result illustrates the localized character of the source term, as expected from resonant, or selective process, of wind-wave generation.

In this figure, solid contours represent positive growth rate, dotted contours represent negative rate (mainly the decay of shoaling swell component), and dashed contours are null values. The discrete "islands" of the calculated growth rate function seem to reflect the localized selective growth character conceptually sketched in Fig. 2. In addition to the two lobes or islands in the input source function as depicted in Fig. 2, experimental data shown in Fig. 3 suggest that a weaker lobe in the direction of the wind. These results seem to render some support to the proposed resonance-type source function, which displays discrete and localized (in the $k$-$\theta$ plane) amplification character of surface waves responding to steady wind forcing.

Presently, the wind-generation source function (of gravity waves) is formulated as a monotonically increasing function with wavenumber. The calculated source term, such as those displayed in Fig. 3, usually show that the net change in the high-frequency or high-wavenumber region is negligibly small. It is generally interpreted that the breaking function also increases with wavenumber at a similar rate as that of the wind-generation function, thus the source and sink terms cancel each other in the short wave region. Recent works on breaking wave observations appear to indicate that the dissipation function is mainly contributed by breaking of "dominant wave" (Banner et al. 2000) rather than a monotonic increase with increasing wavenumber. In other word, the dissipation function is expected to be localized in wavenumber, rather than the present formulation of monotonic increase toward higher wavenumber. From the point view of dynamic balance of the source functions governing the wave evolution process, the independent results of "localized" behavior of breaking waves and wave generation by wind are complementary.

Because surface waves represent the physical roughness of the ocean surface, a clear understanding of the directional property of ocean waves is critical to many applications in air-sea interaction and ocean remote sensing. For example, it has been frequently observed that ocean surface wind stress do not aligned in the direction of wind vector in the presence of ocean swell (e.g., Grachev et al. 2001). The result can be explained by the fact that in the presence of swell, the ocean surface roughness is modulated to align in a direction between the wind vector and the swell propagation direction. Hwang and Shemdin (1988) report the statistic properties of high-frequency waves measured by a laser slope gauge. They show that the primary axis of the 2D probability distribution function (PDF) of small scale ocean surface roughness deviates from the wind direction. The angle of the major axis of the roughness PDF correlates well with the angle between background swell and the wind direction (see Figs. 6 and 7 in Hwang and Shemdin 1988). In the surface wave topography shown in Fig. 1, directional asymmetry is also observed in the two wind-wave systems. It is noted that a mild swell system is present (about 10 s period and 0.3 m wave height). Due to the directional difference between the two wind-wave systems and the swell, the surface waves in the two wave systems are modulated by different degrees. It suggests that a better understanding of air-sea interaction processes, especially the directional properties, relies on a more realistic specification of the directional surface roughness properties.

**4. Summary**

The concept that "wind-generated waves propagate in the wind direction" is deep-rooted and has been assumed either implicitly or explicitly in many aspects of wave research and application. Examples include the design of directional distribution functions and the specification of the wind input term in the wave spectral energy and action density conservation equations. Three-dimensional ocean surface topography observed from photographs, however, indicate that bimodal directional propagation prevails in a young wave field where the phase speed of the dominant wave component is slower than the wind speed. The primary mechanism causing the directional bimodality is probably resonant propagation between surface waves and the wind field. A conceptual model of the source function based on the resonant wind-generation of ocean waves is proposed (Fig. 1). The model suggests that as a consequence of the selective process of resonance generation, a monochromatic wavenumber-dependence of the wind-input function, as traditionally assumed, is unlikely to be correct. A more reasonable function should be bell-shaped, resulting from the characteristic resonance response function of $R \sim \{1-(f/f_r)^m\}^{-n}$. Adding the directionality, the 2D generation function resembles discrete "islands" located at the $(k,\theta)$ coordinates that satisfy the resonant propagation condition.

At a more mature stage, bimodal directional distribution occurs in wave components shorter than the dominant wavelength. The bimodality is due to nonlinear wave-wave interaction. Technology advances in computer, laser ranging, and global positioning have made it possible to acquire high-resolution 3D surface wave topography using spatial measurement systems such as airborne scanning lidar (Hwang et al., 2000a-c). These data provide directional spectra of waves in the decameters length scale. Reliable and accurate directional measurement techniques for small-scale waves (several centimeters to several meters wavelengths) such as those shown in Plates 4 to 6 are still not available for field deployment. Surface waves of a few centimeters to a few meters wavelengths commonly exist at all stages of wind wave generation. Our knowledge of the directional spectral information of these waves in the natural environment is still not very good. It is hoped that the photographs presented in this paper on the bimodal directional propagation of surface waves may stimulate further investigations in this area of research. The directional properties of short waves are critical to our understanding of the ocean surface roughness for ocean remote sensing applications and air-sea interaction studies.

**Acknowledgments**

This work is supported by the Office of Naval Research (Naval Research Laboratory Program Element N62435, "Phase Resolved Nonlinear Transformation of Shoaling Waves." (NRL Contribution PP/7330--00-0080).